\documentclass[preprint,pre,floats,aps,amsmath,amssymb]{revtex4}
\usepackage[percent]{overpic}

\usepackage{bm}
\usepackage[colorlinks=true,allcolors=blue]{hyperref}
\usepackage{graphicx}
\usepackage{cleveref}
\usepackage{mathtools} 
\usepackage{tabularx}
\usepackage{float}
\usepackage{lipsum} % for hrule
\usepackage{marginnote}
\usepackage{subcaption}
\usepackage{booktabs} % For formal tables

\usepackage{color, soul}
\usepackage[normalem]{ulem}
\usepackage{multirow}
\usepackage{amsfonts}

\setstcolor{red}

\newcommand\redsout{\bgroup\markoverwith{\textcolor{red}{\rule[0.5ex]{6pt}{1.5pt}}}\ULon}

\begin{document}

\title{Predictive power-sharing scaling law in double-null L-mode plasmas}
\author{K. Lim$^1$, P. Ricci$^1$, L. Stenger$^1$, B. De Lucca$^1$, G. Durr-Legoupil-Nicoud$^1$, O. F\'evrier$^1$, C. Theiler$^1$ and K. Verhaegh$^2$}
\affiliation{$^1$Ecole Polytechnique Fédérale de Lausanne (EPFL), Swiss Plasma Center (SPC), EPFL SB, Station 13, CH-1015 Lausanne, Switzerland \\$^2$United Kingdom Atomic Energy Authority, Culham Campus, Abingdon, Oxfordshire, OX14 3DB, UK}

\begin{abstract}
The physical mechanisms regulating the power sharing at the outer targets of L-mode double-null (DN) configurations are investigated using nonlinear, flux-driven, three-dimensional two-fluid simulations. Scans of parameters that regulate the turbulent level, such as the plasma resistivity and the magnetic imbalance, reveal that the power asymmetry in DN configurations is determined by the combined effects of diamagnetic drift, turbulence, and geometrical factor. Leveraging these observations, an analytical theory-based scaling law for the power-sharing asymmetry is derived and compared with nonlinear simulations. These comparisons indicate that the scaling law effectively captures the trends observed in simulations. Validation with experimental data from TCV DN discharges demonstrates agreement of the scaling law with the experimental results.
\end{abstract}

\maketitle

\section{Introduction}\label{Sec:1}
Power exhaust is considered among the most pressing issues faced on the way to fusion energy \cite{Donne2019}. To provide an alternative to the conventional, partially detached single-null (SN) H-mode operation envisaged for ITER, while maintaining high core performance and remaining within the material constraints, alternative divertor configurations (ADCs) \cite{Soukhanovskii2017, Wenninger2018,  Reimerdes2020, Militello2021} are being explored in view of next step fusion devices.

Among various ADCs, the double-null (DN) configuration presents several advantages. These include (i) the presence of four strike points that can potentially distribute the heat load on a larger surface, (ii) the high-field side (HFS) that is magnetically disconnected from the turbulent low-field side (LFS), driving a majority of the heat flux toward the outer targets that are located at a larger major radius, offering a significant advantage in terms of power handling \cite{Smick2013, LaBombard2017}, (iii) the quiescent HFS plasma that allows for the installation of RF-antennas in this region, where plasma-surface interactions are reduced \cite{LaBombard2017_2}, (iv) the presence of two X-points that can potentially increase radiative losses through two radiation fronts, offering more control over the plasma conditions \cite{Fevrier2021}, (v) the easier achievement of plasma shapes associated with improved performance, confinement, and plasma beta, compared to single-null (SN) configurations \cite{Kim2023}. For these reasons, a large number of studies in the DN configuration are reported from various tokamaks, including Alcator C-Mod \cite{Brunner2018}, DIII-D \cite{Petrie2001}, EAST \cite{Guo2011}, MAST \cite{Temmerman2011}, START \cite{Morel1999}, and TCV \cite{Fevrier2021}. The DN is also under consideration for implementation in future tokamaks, such as DTT \cite{Contessa2019}, K-DEMO \cite{Im2016}, SPARC \cite{Kuang2020}, and STEP \cite{Wilson2020}.

While a clear advantage of the DN is expected to come from an even distribution of the power load between the outer targets, experiments consistently show a pronounced power-sharing asymmetry between the upper and lower divertor legs, even in balanced DN cases \cite{Marchand1995, Temmerman2010, Brunner2018, Petrie2003, Petrie2006, Liu2012}. Several physical mechanisms are proposed to explain the power load asymmetry, such as cross-field drifts ($E \times B$ and diamagnetic drifts) \cite{Rognlien1999, Cohen1999, Rensink2000, Rubino2020}, ballooning modes \cite{Du2015}, Pfirsch-Schlüter (PS) flows \cite{Schaffer1997, Asakura2004}, flux compression between the two separatrices \cite{Osawa2023}, and different recycling rates at divertor targets \cite{Rensink2000}. However, a clear explanation of the reasons behind this asymmetry remains elusive, and its prediction is limited to the use of relatively simple empirical scaling laws \cite{Petrie1999}.

Experimental observations also show that the magnetic imbalance of the DN configuration significantly affects the heat flux reaching the divertor targets \cite{Petrie2006, Brunner2018}. When unbalanced, the DN configuration presents two separatrices associated with the upper and lower X-points, their distance at the outboard midplane being denoted as $\delta R$. Determining the optimal value of $\delta R$ is of crucial importance for improving the heat exhaust performance of the DN configuration. For example, an increased $\delta R$ approaching single null (SN) configurations leads to an enhanced power load at the inner targets \cite{Brunner2018}. We also note that, in addition to the heat load sharing, the magnetic imbalance in DN has an impact on various factors, such as the L-H transition power threshold \citep{Maggi2014}, detachment conditions \cite{Fevrier2021}, and the density limit \cite{Petrie1999}.

Several fluid models, such as GBS \cite{Beadle2020}, SOLPS-ITER \cite{Ahomantila2021, Osawa2023}, SOLEDGE-2D \cite{Innocente2021}, and UEDGE \cite{Petrie1999, Porter2010}, have previously investigated the power load in DN configurations and its possible advantages in power handling compared to SN configurations. In the present study, we leverage previous turbulent simulations in DN scenarios \cite{Beadle2020} carried out by using the GBS code, which reveals the physical mechanisms that determine the density decay length in the SOL region of the DN configuration. We extend this analysis to derive an analytical expression for the pressure gradient length $L_p$. We then examine the effects of magnetic imbalance, plasma resistivity, and the direction of the toroidal magnetic field on heat load asymmetry. We derive a first-principle analytical scaling law to predict power-sharing imbalance on the outer targets in DN configuration. The theoretical scaling law for the heat asymmetry is finally compared with a set of nonlinear GBS simulations, and validated against experimental data from discharges carried out in the Tokamak à Configuration Variable (TCV) \cite{Reimerdes2022}.

The remainder of the paper is organized as follows. Section \ref{Sec2} introduces the GBS model, detailing the setup parameters for a series of nonlinear simulations, as well as a description of the reference DN magnetic configuration. In Section \ref{Sec3}, we present the results obtained from the nonlinear GBS simulations, with focus on the heat flux asymmetry between the four divertor legs. Section \ref{Sec4} develops an analytical scaling law to predict power-sharing imbalances between the upper and lower divertor targets, which is verified against nonlinear GBS simulations. Section \ref{Sec5} provides a validation of the scaling law with experimental data from TCV DN discharges. Finally, the conclusions are drawn in Section \ref{Sec6}.

\section{Numerical model}\label{Sec2}
The simulation of plasma turbulence in the tokamak boundary is often based on the drift-reduced Braginskii model \cite{Zeiler1997}. The use of this model is justified by the characteristic time scales of plasma turbulence, longer than $1/\Omega_{ci}$, with $\Omega_{ci}=eB/m_i$ the ion cyclotron frequency, and the parallel length scales longer than $\lambda_e$, the electron mean-free path. 

The three-dimensional drift-reduced Braginskii equations are implemented in the GBS code that we use here to explore plasma turbulence in the DN configuration \cite{Ricci2012, Giacomin2021}. We note that, in recent years, the plasma model implemented in GBS has been developed to include the interaction with neutrals, to extend the simulation domain encompassing the overall plasma volume, and to allow for a non-Boussinesq treatment of the polarization drift \cite{Giacomin2021}. The GBS implementation of the drift-reduced Braginskii equations uses a coordinate system that is independent of the magnetic field, enabling simulations in arbitrary magnetic configurations, such as snowflake \cite{Giacomin2020}, negative triangularity \cite{Riva2017, Lim2023}, DN \cite{Beadle2020}, as well as non-axisymmetric \cite{Coelho2022} configurations.

In the present study, we restrict ourselves to the axisymmetric magnetic field in the electrostatic limit and, for simplicity, we avoid simulating the interaction between plasma and neutrals. Consequently, the GBS model equations can be recast as follows:
\begin{widetext}
\begin{align}
     \frac{\partial n}{\partial t} = -\frac{\rho_*^{-1}}{B}[\phi, n] + \frac{2}{B}\bigg[C(p_e)-nC(\phi)\bigg]-\nabla_\parallel (nv_{\parallel e}) + D_n \nabla^2_\perp n + s_n,
\label{GBS_density}
\end{align}

\begin{align}
    \frac{\partial \Omega}{\partial t} = -\frac{\rho_*^{-1}}{B} \nabla \cdot [\phi, \omega] -\nabla \cdot \big(v_{\parallel i} \nabla_\parallel\omega\big) + B^2 \nabla_\parallel j_\parallel + 2B C(p_e + \tau p_i) + \frac{B}{3}C(G_i) + D_\Omega \nabla_\perp^2 \Omega,
\label{GBS_vorticity}
\end{align}
\begin{align}
    \frac{\partial v_{\parallel i}}{\partial t} = -\frac{\rho_*^{-1}}{B}[\phi, v_{\parallel i}]-v_{\parallel i} \nabla_\parallel v_{\parallel i} - \frac{1}{n}\nabla_\parallel (p_e + \tau p_i) -\frac{2}{3n}\nabla_\parallel G_i + D_{v_{\parallel i}}\nabla_\perp^2 v_{\parallel i},
\end{align}
\begin{align}
     \frac{\partial v_{\parallel e}}{\partial t} &= -\frac{\rho_*^{-1}}{B}[\phi, v_{\parallel, e}] - v_{\parallel e}\nabla_\parallel v_{\parallel e} + \frac{m_i}{m_e}\Bigg(\nu j_\parallel + \nabla_\parallel \phi -\frac{1}{n}\nabla_\parallel p_e-0.71 \nabla_\parallel T_e -\frac{2}{3n}\nabla_\parallel G_e \Bigg) \nonumber \\
    & + D_{v_{\parallel e}}\nabla_\perp^2 v_{\parallel e}, 
    \label{GBS_ve}
\end{align}
\begin{align}
     \frac{\partial T_i}{\partial t} &=-\frac{\rho_*^{-1}}{B}[\phi, T_i] - v_{\parallel i}\nabla_\parallel T_i + \frac{4}{3}\frac{T_i}{B}\Bigg[C(T_e) + \frac{T_e}{n}C(n)-C(\phi) \Bigg] -\frac{10}{3}\tau \frac{T_i}{B}C(T_i) \nonumber \\
     &+ \frac{2}{3}T_i\Bigg[(v_{\parallel i}-v_{\parallel e})\frac{\nabla_\parallel n}{n}-T_i \nabla_\parallel v_{\parallel e}\Bigg] + 2.61\nu n (T_e -\tau T_i)+ \nabla_\parallel( \chi_{\parallel i}\nabla_\parallel T_i) \nonumber\\
     &+ D_{T_i}\nabla_\perp^2 T_i + s_{T_i},
\end{align}
\begin{align}
     \frac{\partial T_e}{\partial t} &= -\frac{\rho_*^{-1}}{B}[\phi, T_e] - v_{\parallel e}\nabla_\parallel T_e + \frac{2}{3}T_e\Bigg[0.71 \frac{\nabla_\parallel j_\parallel}{n}-\nabla_\parallel v_{\parallel e}\Bigg] -2.61\nu n (T_e-\tau T_i)
    \nonumber \\
    &+ \frac{4}{3}\frac{T_e}{B}\Bigg[\frac{7}{2}C(T_e)+\frac{T_e}{n}C(n)-C(\phi) \Bigg] + \nabla_\parallel (\chi_{\parallel e}\nabla_\parallel T_e) + D_{T_e}\nabla_\perp^2 T_e + s_{T_e}, \label{GBS_electron_temperature}
\end{align}
\end{widetext}
these equations being coupled with the Poisson equation that avoids the Boussinesq approximation
\begin{align}
    \nabla \cdot \bigg( n\nabla_\perp \phi \bigg) = \Omega - \tau \nabla_\perp^2 p_i,
\label{GBS_Poisson}
\end{align}
where $\Omega=\nabla \cdot \omega = \nabla \cdot (n\nabla_\perp \phi + \tau \nabla_\perp p_i)$ is the scalar vorticity.

In Eqs.\,(\ref{GBS_density}--\ref{GBS_Poisson}), the plasma variables are normalized to reference values. Specifically, the plasma density, $n$, the ion and electron temperatures, $T_{i}$ and $T_e$, the ion and electron parallel velocities, $v_{\parallel i}$ and $v_{\parallel e}$, and the electric potential $\phi$ are normalized to the reference values $n_0, T_{i0}, T_{e0}, c_{s0}=\sqrt{T_{e0}/m_i}$ and $T_{e0}/e$, respectively. The perpendicular lengths are normalized to the ion sound Larmor radius $\rho_{s0}=c_{s0}/\Omega_{ci}$, while parallel lengths to the tokamak major radius $R_0$. Time is normalized to $R_0/c_{s0}$.

The following dimensionless parameters that govern plasma dynamics appear in Eqs.\,(\ref{GBS_density}--\ref{GBS_Poisson}): the normalized ion Larmor radius $\rho_*=\rho_{s0}/R_0$, the ratio of ion to electron temperature $\tau=T_{i0}/T_{e0}$, the normalized ion and electron viscosity $\eta_{0i}=0.96 n T_i \tau_i$ and $\eta_{0e}=0.73 n T_e \tau_e$, respectively, where $\tau_{e,i}$ represents the electron and ion collision times, and the normalized ion and electron parallel thermal conductivity, 
\begin{align}
    \chi_{\parallel i} = \bigg(\frac{1.94}{\sqrt{2\pi}}\sqrt{m_i} \frac{(4\pi \epsilon_0)^2}{e^4} \frac{c_{s0}}{R_0}\frac{T_{e0}^{3/2}\tau^{5/2}}{\lambda n_0} \bigg) T_i^{5/2}
\end{align}
and 
\begin{align}
    \chi_{\parallel e} = \bigg(\frac{1.58}{\sqrt{2\pi}}\frac{m_i}{\sqrt{m_e}}\frac{(4\pi \epsilon_0)^2}{e^4} \frac{c_{s0}}{R_0}\frac{T_{e0}^{3/2}}{\lambda n_0} \bigg) T_e^{5/2},
\end{align}
The normalized Spitzer resistivity is defined as $\nu=e^2n_0R_0/(m_ic_{s0}\sigma_\parallel)=\nu_0 T_e^{-3/2}$, with
\begin{align}
    \sigma_\parallel = \Bigg(1.96\frac{n_0 e^2 \tau_e}{m_e}\Bigg)n = \Bigg(\frac{5.88}{4\sqrt{2\pi}} \frac{(4\pi \epsilon_0)^2}{e^2}\frac{T_{e0}^{3/2}}{\lambda \sqrt{m_e}}\Bigg)T_e^{3/2}
\end{align}
and
\begin{align}
    \nu_0 =\frac{4\sqrt{2\pi}}{5.88} \frac{e^4}{(4\pi \epsilon_0)^2}\frac{\sqrt{m_e}R_0 n_0 \lambda}{m_i c_{s0}T_{e0}^{3/2}},
\label{Collisionality}
\end{align}
where $\lambda$ is the Coulomb logarithm.

The differential geometrical operators described in Eqs.\,(\ref{GBS_density}--\ref{GBS_Poisson}) are defined as follows:
\begin{align}
    [\phi,f] &= \bm{b} \cdot (\nabla \phi \times \nabla f), \label{GBS_operator1}\\
    \mathcal{C}(f) &= \frac{B}{2}\bigg( \nabla \times \frac{\bm{b}}{B}\bigg) \cdot \nabla f, \label{GBS_operator2} \\
    \nabla_\parallel f &= \bm{b} \cdot \nabla f, \\
    \nabla_\perp^2 f &=\nabla \cdot \big[(\bm{b}\times \nabla f)\times \bm{b}\big].\label{GBS_operator4}
\end{align}
These correspond to the $E \times B$ convective term, the curvature operator, the parallel gradient, and the perpendicular Laplacian of a scalar function $f$. To compute these differential operators, the $(R, \varphi, Z)$ non-field-aligned cylindrical coordinates are used, where $R$ represents the radial distance from the magnetic axis, $Z$ is the vertical coordinate, and $\varphi$ is the toroidal angle. An axisymmetric magnetic field is considered, expressed as $\bm{B}=RB_\varphi\nabla \varphi + \nabla \varphi \times \nabla \psi$ where $\psi(R,Z)$ is the poloidal magnetic flux. It is important to note that the operators defined in Eqs.\,(\ref{GBS_operator1}-\ref{GBS_operator4}) depend on the sign of the normalized magnetic field $\bm{b}=\bm{B}/B$. A positive sign indicates that the magnetic drifts are directed toward the lower X-points (favorable direction), whereas a negative sign indicates the opposite direction (unfavorable).  

The gyroviscous terms are defined as 
\begin{eqnarray}
    G_i=-\eta_{0i}\Bigg[2\nabla_\parallel v_{\parallel i} + \frac{1}{B}C(\phi) + \frac{1}{enB}C(p_i) \Bigg]
\end{eqnarray}
and
\begin{eqnarray}
    G_e=-\eta_{0e}\Bigg[2\nabla_\parallel v_{\parallel e} + \frac{1}{B}C(\phi)-\frac{1}{enB}C(p_e) \Bigg].
\end{eqnarray}
The artificial diffusion terms $D_f\nabla_\perp^2 f$ are introduced to the right-hand side of Eqs.\,(\ref{GBS_density}--\ref{GBS_Poisson}) to improve the numerical stability of the simulations.

\iffalse
The poloidal magnetic flux, $\psi(R,Z)$, is generated using a Gaussian-like current at the magnetic axis along with additional current filaments (C1-C8) located outside the simulation domain. One example of the magnetic equilibrium used in the present study is displayed in Fig.\,\ref{Fig:DN_coils} for a balanced DN configuration. While the positions of these filaments are kept constant throughout this study, the amplitude of each coil is adjusted to create both balanced and unbalanced DN configurations. The configuration we consider is characterized by elongation $\kappa \simeq 1.7$ and we impose, for the sake of simplicity, vanishing triangularity, $\delta=0$, to focus on the effect of magnetic imbalance of the DN configurations.

\begin{figure}
\begin{center}
\includegraphics[width=0.45\textwidth]{Figures/DN_coils.pdf}
\caption{Magnetic equilibrium profile used for the nonlinear GBS double-null simulations (balanced configuration). The black crosses indicate the positions of the current-carrying coils that generate the magnetic field. The red solid line is the separatrix.}
\label{Fig:DN_coils}
\end{center}
\end{figure}
\fi

The presence of a toroidally uniform density and temperature source inside the last closed flux surface (LCFS) mimics the ionization process and Ohmic heating within the core. They assume the following expressions:
\begin{align}
    s_n &= s_{n0}\exp{\Bigg( -\frac{[\psi(R,Z)-\psi_n]^2}{\Delta_n^2}\Bigg)}\label{density_source}
\end{align}
and
\begin{align}
    s_T &= \frac{s_{T0}}{2}\Bigg[\tanh\Bigg(-\frac{\psi(R,Z)-\psi_T}{\Delta_T}\Bigg)+1\Bigg], \label{temperature_source}
\end{align}
where $\psi_n$ and  $\psi_T$ are flux function values inside the LCFS, while $\Delta_n$ and $\Delta_T$ determine the radial width of the source terms. 

The presence of a strong electric field in the magnetic pre-sheath violates the assumptions underlying the drift approximation. As a result, the boundary conditions that satisfy the Bohm-Chodura criterion are implemented at the magnetic pre-sheath entrance \cite{Loizu2012}, where the gradients of density and electrostatic potential in directions tangent to the wall are neglected:
\begin{align}
    v_{\parallel i}&=\pm \sqrt{T_e+\tau T_i}, \\
    v_{\parallel e}&=\pm\sqrt{T_e + \tau T_i}\exp{\Bigg( \Lambda - \frac{\phi}{T_e}\Bigg)}, \\
    \partial_Z n &= \mp \frac{n}{\sqrt{T_e+\tau T_i}}\partial_Z v_{\parallel i}, \\
    \partial_Z \phi &= \mp \frac{T_e}{\sqrt{T_e + \tau T_i}}\partial_Z v_{\parallel i},\\
    \partial_Z T_e &= \partial_Z T_i = 0, \\
    \omega &= -\frac{T_e}{T_e + \tau T_i}\bigg[(\partial_Z v_{\parallel i})^2 \pm \sqrt{T_e + \tau T_i} \partial^2_Z v_{\parallel i}\bigg].
\end{align}
Here, the $\pm$ sign indicates whether the magnetic field line enters (top sign) or leaves (bottom sign) the wall, and $\Lambda = \log \sqrt{m_i/(2\pi m_e)}\simeq 3$. For the left and right walls within the simulation box, the electric potential is defined as $\phi=\Lambda T_e$, and the derivatives normal to the wall are set to vanish for all other quantities. 

\section{Overview of simulation results}\label{Sec3}
In this section, we first discuss the setup of the nonlinear GBS simulations we present. We then analyze the numerical results focusing on the mechanisms that set the SOL pressure decay length and the power asymmetry in both balanced and unbalanced configurations. These are the key elements used in the derivation of the scaling law for the power sharing.

\subsection{Simulation setup}
The poloidal magnetic flux, $\psi(R,Z)$, is generated using a Gaussian-like current at the magnetic axis along with additional current wires (C1-C8) located outside the simulation domain. While the positions of these wires are kept constant throughout this study, the amplitude of the current is adjusted to create both balanced and unbalanced DN configurations, as shown in Fig.\,\ref{Fig:DN_coils}. The configuration we consider is characterized by elongation $\kappa \simeq 1.7$ and we impose, for the sake of simplicity, vanishing triangularity, $\delta=0$, to focus on the effect of magnetic imbalance of the DN configurations. Furthermore, we position artificially designed targets, Lower Outer (LO), Upper Outer (UO), Lower Inner (LI) and Upper Inner (UI), to assess the total heat flux reaching the divertor targets. These targets are placed slight away from the boundary of the simulation domain to avoid the effect of the strong gradients that often appear in this region and that challenge the numerical evaluation of the fluxes, as illustrated in Fig. \ref{Fig:DN_coils}.

In the unbalanced DN, the upper and lower X-points are located on different separatrices. In the LSN configuration, the primary separatrix that encompasses the core region is connected to the bottom wall, whereas it is connected to the upper wall in the USN configuration. To evaluate the magnetic imbalance, we introduce the inter-separatrix distance, $\delta R=R_L - R_U$, which is the radial distance at the midplane between the separatrices connected to the lower and upper X-points. Negative values represent the LSN configuration, while positive values indicate the USN configuration. Large $\lvert\delta R\rvert$ values yield a secondary separatrix that has a negligible impact on plasma dynamics, thereby the DN approaches a SN configuration in this case.

\begin{figure*}
\begin{center}	
\includegraphics[width=\textwidth]{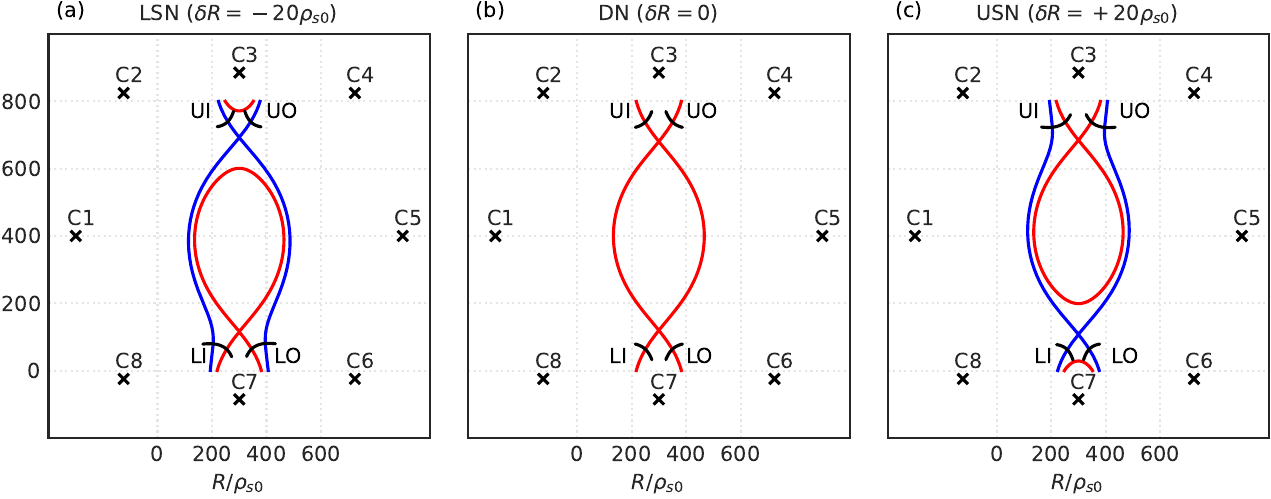}
\caption{Three magnetic equilibria, (a) LSN, (b) DN, and (c) USN, used for the nonlinear GBS simulations. The black crosses indicate the positions of the current-carrying coils that generate the magnetic field, while the black dot represents the core current at the magnetic axis. The main separatrix is traced by a red solid line, and the secondary separatrix by a blue solid line. Four solid lines are positioned in front of each divertor target, denoted by Lower Outer (LO), Upper Outer (UO), Lower Inner (LI), and Upper Inner (UI), where the target heat flux is evaluated. Plasma elongation is set to $\kappa \simeq 1.7$, triangularity to $\delta=0$.}
\label{Fig:DN_coils}
\end{center}
\end{figure*}

\iffalse %(KL)
\begin{figure*}
\begin{center}	
\includegraphics[width=\textwidth]{Figures/Equil_DN.pdf}
\caption{Three magnetic equilibria, i.e. LSN (left), DN (center), and USN (right), used for the nonlinear GBS simulations. The red solid line represents the main separatrix and the blue solid line indicates the secondary separatrix. Plasma elongation is set to $\kappa \simeq 1.7$, triangularity to $\delta=0$.}
\label{Fig:DN_LSN_USN}
\end{center}
\end{figure*}
\fi
For all GBS simulations presented in this study, we use a numerical grid defined by $(N_R, N_Z, N_\varphi) = (240, 320, 80)$, with a time step $\Delta t=10^{-5}R_0/c_{s0}$. The size of the simulation domain is set to $L_R=600 \rho_{s0}, L_Z=800 \rho_{s0}$ and $ \rho_*=\rho_{s0}/R_0=1/700$. These correspond to a tokamak that is, approximately, one-third the size of TCV \cite{Reimerdes2022}. For the radial width of the source terms, we use $\Delta_n=1000$ and $\Delta_T=900$, along with flux function values $\psi_n=120$ and $\psi_T=110$ in Eqs.\,(\ref{density_source}--\ref{temperature_source}). The same amplitude, $s_{T0}$, is used for both the ion and electron temperature sources. We consider a ratio $m_i/m_e=200$, which reduces the computational cost of our simulations and allows us to carry out a parameter scan. The small mass ratio is not expected to have a significant impact on the results of nonlinear simulations, since in the case of L-mode plasmas simulated here, the plasma resistivity dominates over inertial effects when $\nu_0>(m_e/m_i)\gamma$ with $\gamma$ denoting the growth rate. To facilitate the analysis, the plasma parameters $\tau=T_{i0}/T_{e0}=1$,  $\eta_{0e}=\eta_{0i}=1$ and $ \chi_{\parallel e}=\chi_{\parallel i}=1$, are kept constant throughout the study. The safety factor is adjusted such that $q_0 \simeq 1$ at the magnetic axis and $q_{95}\simeq 4$ at the tokamak edge. Additionally, the reference toroidal magnetic field $B_T$ is set for the ion-$\nabla B$ drift being away from the lower X-point. To explore the effects of different turbulent regimes and magnetic imbalances on heat distribution in DN configurations, we vary the plasma resistivity $\nu_0=\{0.1, 0.3, 1.0\}$, the amplitude of the input heating power $s_{T0}=\{ 0.075, 0.15, 0.3\}$, the magnetic imbalance $\delta R=\{-40\rho_{s0}, -30\rho_{s0}, -20\rho_{s0}, 20\rho_{s0}, 30\rho_{s0}, 40\rho_{s0}\}$ across both LSN and USN configurations. All simulations are carried out in the reverse $B_T$ configuration, and a forward $B_T$ simulation is also considered for the case with $\nu_0=0.3$ and $s_{T0}=0.15$. The ranges for the plasma resistivity and the heating source amplitude are selected to ensure that plasma turbulence is dominated by resistive ballooning modes (RBMs) in our simulations, which are destabilized by the magnetic field curvature and the plasma pressure gradient at sufficiently high plasma resistivity \cite{Zeiler1997}. This regime is consistent with the tokamak L-mode operation \cite{Giacomin2020}.

\begin{figure}[H]
\begin{center}	
\begin{overpic}[width=\textwidth]{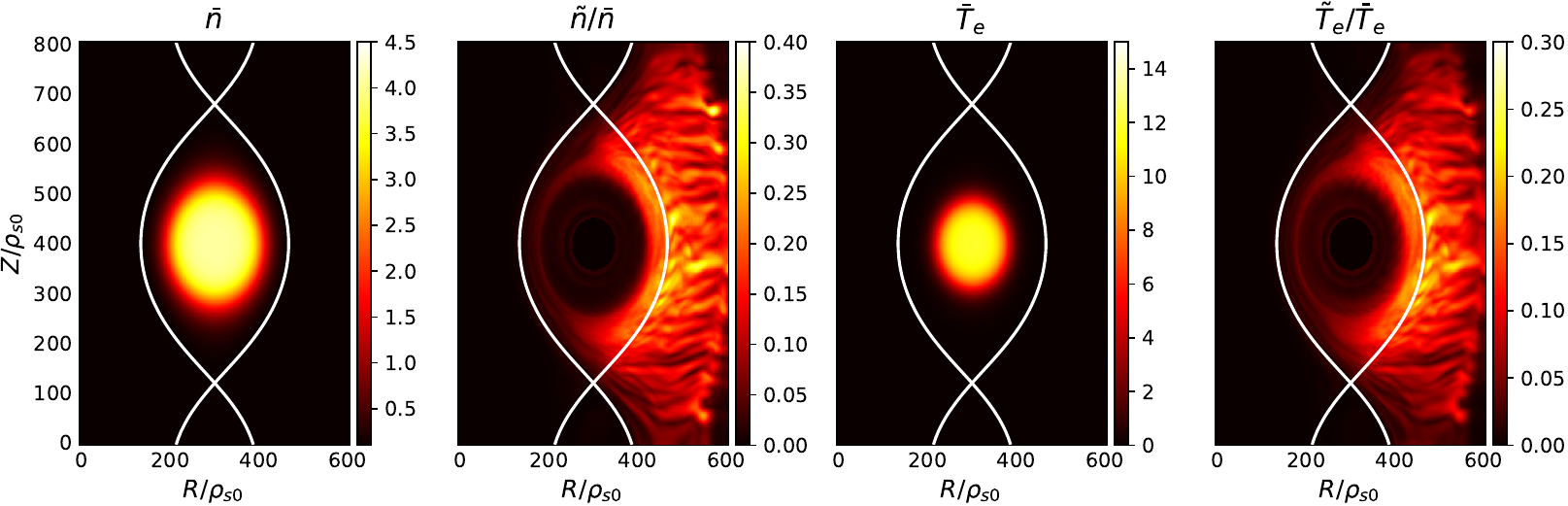}
  \put(5,30.5){(a)}  
  \put(29,30.5){(b)}  
  \put(53,30.5){(c)} 
  \put(77,30.5){(d)}   
\end{overpic}
\caption{Two-dimensional snapshots of (a) averaged density (b) normalized density fluctuations (c) averaged electron temperature (d) normalized electron temperature fluctuations for the simulations with $\nu_0=0.3$ and $s_{T0}=0.15$. }
\label{Fig:2D_snapshot}
\end{center}
\end{figure}
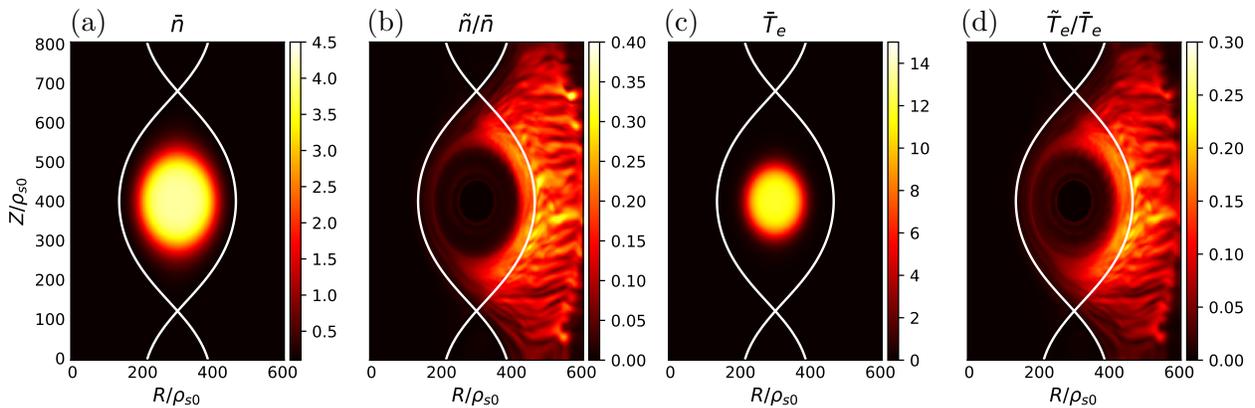

After a transient phase, the simulations reach a quasi-steady state where the input heating power, perpendicular transport, and losses at the vessel wall balance each other, causing all quantities to fluctuate around constant values. Once the quasi-steady state is attained, all quantities are averaged toroidally and over a time window of $10t_0$. Throughout this paper, the time- and toroidally-averaged quantities are represented with an overline, while fluctuating quantities are expressed with a tilde, i.e. $f=\bar{f}+\tilde{f}$. Figure \ref{Fig:2D_snapshot} presents a typical two-dimensional snapshot illustrating both the equilibrium and relative fluctuating components of the density and electron temperature. The fluctuating components reveal that strong turbulence develops at the LFS in the edge and near SOL regions, which propagates to the far SOL region because of the presence of blobs, whereas the HFS region remains quiescent as it is magnetically disconnected from the LFS. 

\subsection{Estimate of the pressure gradient length}
\iffalse
Figure \ref{Fig:2D_fluct} describes the poloidal section of fluctuating density terms in different magnetic configurations. The strongly developed fluctuating structure at the LFS indicates the presence of interchange-type instabilities, i.e. resistive ballooning modes (RBMs), in the present simulations. The unbalanced configurations considered in Fig.\,\ref{Fig:2D_fluct} have $\delta R_{\textrm{sep}}=\pm 20 \rho_s$. The heat flux arriving at the target linked to the separatrix is facilitated when the radial size of the turbulence structure is larger than $\delta R_{\textrm{sep}}$. If $\delta R_{\textrm{sep}} k_r > 1$, the majority of the heat flux would flow onto the targets linked to the primary separatrix.

\begin{figure}[H]
\centering
\includegraphics[width=\textwidth]{Figures/Fluct_2D_density.png}
\caption{2D poloidal section of fluctuating density terms. The line where heat fluxes are measured illustrated in different color sets. For unbalanced configurations, the values of $\delta R_{\textrm{sep}}=\pm 20 \rho_s$.}
\label{Fig:2D_fluct}
\end{figure}
\fi

\begin{figure}[H]
\begin{center}
\begin{overpic}[width=0.7\textwidth]{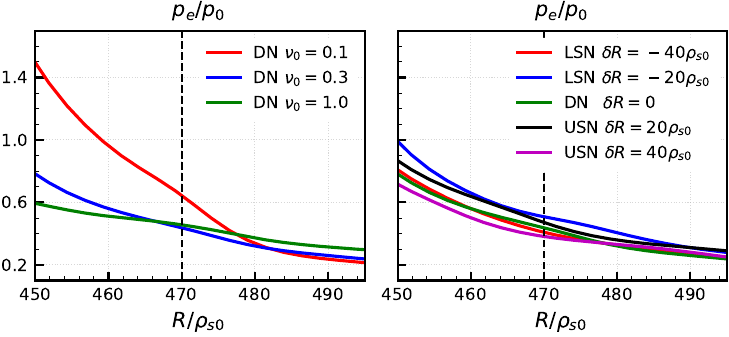}
  \put(6,38.5){(a)}  
  \put(56,38.5){(b)}  
\end{overpic}
\caption{Radial OMP profiles of electron pressure across various resistivities and magnetic configurations: On the left, a scan in plasma resistivity $\nu_0$ is conducted within a balanced DN configuration; on the right, a scan in magnetic imbalance $\delta R$ is performed with $\nu_0=0.3$ constant. The position of the separatrix is represented with the dashed line.}
\label{Fig:OMP}
\end{center}
\end{figure}

Figure \ref{Fig:OMP}\textcolor{blue}{a} shows the radial profile of electron pressure at the outer midplane (OMP), considering a scan in plasma resistivity $\nu_0$ in a balanced DN configuration. An increase of the turbulent transport is observed with resistivity, yielding a weaker pressure gradient at high resistivity \cite{Mosetto2013, Giacomin2020}. We also consider a scan of the magnetic imbalance in Fig.\,\ref{Fig:OMP}\textcolor{blue}{b} and, interestingly, the radial $p_e$ profile at the OMP is observed not to be strongly affected by variations in $\delta R$ values across both LSN and USN configurations. The minor difference of radial $p_e$ profiles between the LSN and USN can be attributed to the direction of the $\nabla B$-drift, which is directed towards the X-points in the case of LSN (favorable) and away in the case of USN (unfavorable). Indeed, while the shape of the separatrix (e.g., triangularity) is observed to affect the turbulence in the boundary \cite{Lim2023}, the introduction of a secondary separatrix in an unbalanced DN configuration appears not to have a significant impact in our simulations. On the other hand, the particle and power exhaust at the targets are influenced by the variations in $\delta R$, as shown by our simulations and previous studies \cite{Petrie1999, Temmerman2010}.

The pressure gradient length in the near SOL, denoted by $L_{p}=-p_e/\nabla p_e$, is found experimentally to correlate with the power fall-off decay length $\lambda_q$ at the outer targets \cite{Silvagni2020}. Based on this correlation, theoretical scaling laws relating tokamak operational parameters to $\lambda_q$ in L-mode plasmas with diverted SN configurations are derived from a first-principle approach in Refs.\,\cite{Giacomin2021_2, Lim2023}. These scaling laws are validated against both nonlinear GBS simulations and multi-machine datasets, demonstrating their reliability in predicting the SOL width based on various engineering parameters \cite{Giacomin2021_2, Lim2023}. Leveraging these studies, as well as the investigations of the density decay-length $L_n$ outlined in Ref.\,\cite{Beadle2020}, an analytical estimate of $L_p$ in DN configurations can be derived. 

The derivation is based on a gradient removal theory \cite{Ricci2008, Ricci2013}, where the local flattening of the plasma pressure profile provides the main mechanism for the saturation of the growth of linear instabilities driving turbulence. Then, the value of $L_p$ is determined through a balance between perpendicular turbulent transport and parallel losses at the end of the magnetic field lines. The derivation of the $L_p$ scaling is identical to the one for the SN configuration, which is detailed in Ref.\,\cite{Lim2023}, and is consistent with the results shown in Fig.\,\ref{Fig:OMP}, highlighting that the introduction of a secondary separatrix does not appear to significantly modify the pressure gradient length $L_p$ in the near SOL. The analytical expression for the $L_p$ estimate in DN configurations can be written as:
\begin{align}
    L_{p} \sim \mathcal{C}(\kappa, \delta, q) \bigg[ \rho_* (\nu_0 \bar{n} q^2)^2 \bigg(\frac{L_\chi \bar{p}_e}{S_p}\bigg)^{4} \bigg]^{1/3},
\label{Lp_analytical}
\end{align}
where $q$ is the safety factor at the tokamak edge, $L_\chi\simeq \pi a (0.45 + 0.55\kappa)$ is an approximation of the poloidal length of the LCFS for an elongation near $\kappa=1$, and $S_p$ represents the volume integrated heat source. The curvature coefficient is defined with values at the outer midplane, since RBMs are mostly destabilized at the LFS, is defined as follows
\begin{align}
    \mathcal{C}(\kappa, \delta, q) = 1-\frac{\kappa-1}{\kappa+1}\frac{3q}{q+2} + \frac{\delta q}{1+q} + \frac{(\kappa-1)^2(5q-2)}{2(\kappa+1)^2(q+2)} + \frac{\delta^2}{16}\frac{7q-1}{1+q}.
    \label{Curvature_operator}
\end{align}

The expression for $L_p$, consistently with the results shown in Fig.\,\ref{Fig:OMP}, highlights that the introduction of a secondary separatrix does not appear to significantly modify the pressure gradient length $L_p$ in the near SOL.

\begin{figure}[H]
\begin{center}
\includegraphics[width=\textwidth]{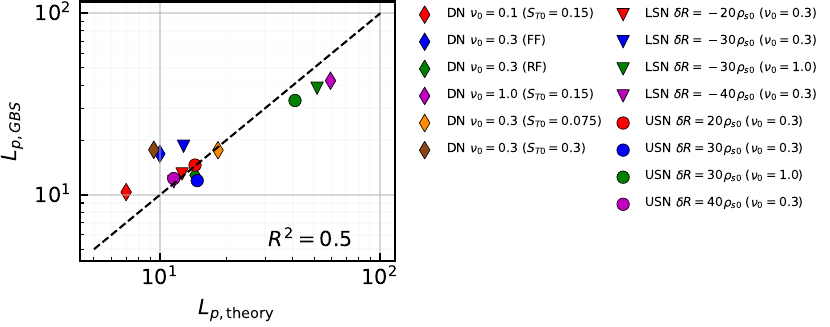}
\caption{Comparison of the pressure gradient length $L_p$ between the analytical scaling law in Eq.\,(\ref{Lp_analytical}) and the value of $L_p$ obtained from nonlinear GBS simulation. A parametric scan for plasma resistivity $\nu_0$ and heating power $S_{T0}$, and the direction of the magnetic field $B_T$, is carried out for different DN configurations. Here,  `FF' denotes the forward field (i.e., ion-$\nabla B$ drift downwards), while 'RF' represents the reverse field.}
\label{Fig:Lp_comparison}
\end{center}
\end{figure}

A comparison between the analytical $L_p$ estimate derived in Eq.\,(\ref{Lp_analytical}) and the values obtained from nonlinear GBS simulations is presented in Fig.\,\ref{Fig:Lp_comparison}. The analytical scaling law captures the general trend with an $R^2$-score of 0.5. In agreement with the analytical expression derived in Eq.\,(\ref{Lp_analytical}), $L_p$ increases proportionally to the value of plasma resistivity $\nu_0$, while it is inversely proportional to the input heating power $S_{T0}$. This is in agreement with the fact that high resistivity and reduced heating power enhance the level of turbulence fluctuations and related transport \cite{Giacomin2020}. For example, in the case of an enhanced turbulence regime at $\nu_0=1$, the total heat flux crossing the separatrix shows a 25\% increase when compared to the reduced turbulence regime at $\nu_0=0.1$. This leads to the flattening of the pressure gradient and, as a result, yielding a larger value of $L_p=-p_e/ \nabla p_e$. Moreover, the value of $L_p$ remains relatively unaffected in different magnetic configurations (DN, LSN, and USN) when $\nu_0$ is held constant, consistently with Fig.\,\ref{Fig:OMP}. Instead, it is mainly affected by the intensity of background plasma turbulence.

\subsection{Power load asymmetry on divertor legs}
Imbalanced heat distribution between different divertor legs is observed in present tokamaks \cite{Brunner2018, Petrie2001, Guo2011, Temmerman2011, Morel1999, Fevrier2021}. While the asymmetry between inner and outer targets in SN configurations is often attributed to the radial and poloidal components of the $E \times B$ drift terms \cite{ Rognlien1999, Rognlien1999_2, Chankin2015, Christen2017}, the heat distribution asymmetry between upper and lower targets in DN configurations results from a complex interplay of various factors. These include particle drifts \cite{Stangeby1996, Rognlien1999, Liu2012, Chankin2015}, interchange-like instabilities that are particularly developed in the bad-curvature region \cite{LaBombard2017}, resistive ballooning modes (RBMs) \cite{Du2015}, Pfirsch-Schlüter (PS) flows \cite{Asakura2000, Asakura2004}, and differences in recycling rates at the divertor targets \cite{Rensink2000}. 

In L-mode plasmas within the sheath-limited regime, characterized by a small temperature gradient between the upstream and the divertor targets, the conductive heat flux can be neglected compared to the convective term. This is supported by our nonlinear DN simulations, which indicate a weak parallel temperature gradient within the sheath-limited regime. Consequently, we consider only the convective parallel heat flux arriving at each divertor target. Assuming $T_e=T_i$, we express the parallel heat flux as $q_\parallel =(5/2)\Gamma_{\parallel e}T_e$, where $\Gamma_{\parallel e}=nv_{\parallel e}$ is the parallel electron particle flux.

\begin{figure}[H]
\begin{center}	
\includegraphics[width=\textwidth]{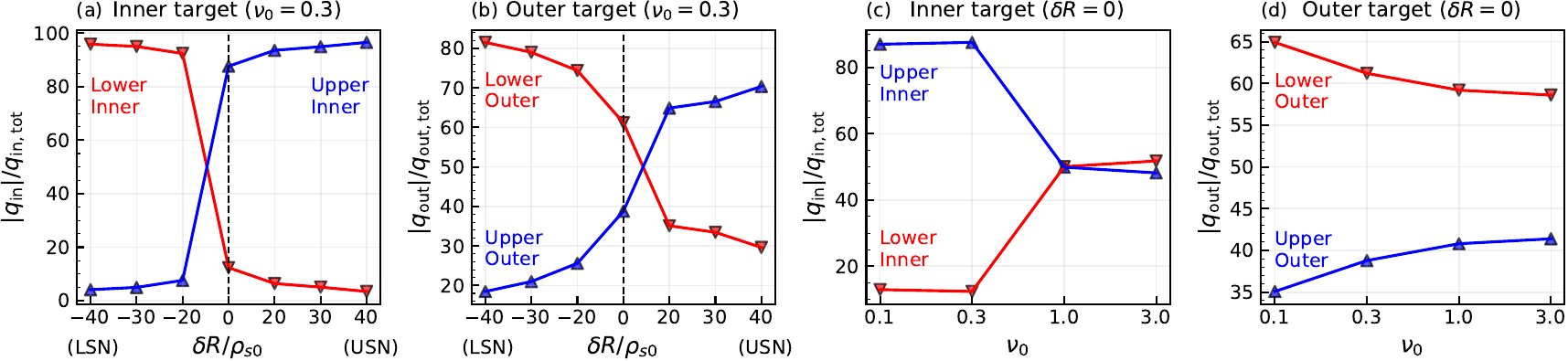}
\caption{Asymmetry of the heat flux reaching the upper and lower divertor targets. We vary the magnetic imbalance while keeping the plasma resistivity constant at $\nu_0=0.3$ [(a) and (b)] and  we vary the plasma resistivity in the balanced DN [(c) and (d)]. Inner heat fluxes are normalized to the total heat flux reaching the HFS targets, while the outer heat fluxes are normalized to the total heat flux directed to the LFS.}
\label{Fig:heat_asymmetry}
\end{center}
\end{figure}

Figure \ref{Fig:heat_asymmetry} illustrates the asymmetry of the total heat flux reaching the four divertor targets that are shown in Fig.\,\ref{Fig:DN_coils}. In this analysis, we vary (i) the magnetic imbalance while keeping a constant plasma resistivity at $\nu_0=0.3$, and (ii) the plasma resistivity in the balanced DN configuration where $\delta R=0$. We note that the heat flux arriving at the inner targets is significantly smaller, below 10\%, compared to that directed towards the outer targets, as the HFS is magnetically disconnected from the LFS \cite{Smick2013, LaBombard2017}. The effect of magnetic imbalance is particularly pronounced at the HFS due to the absence of turbulence, as shown in Fig.\,\ref{Fig:heat_asymmetry}\textcolor{blue}{a}. Specifically, over 80\% of HFS heat flux is directed toward the upper inner target in the balanced DN configuration $(\delta R=0)$. In the case of strongly imbalanced LSN ($\delta R<0$), the majority of the HFS heat flux is directed along the primary separatrix, yielding over 90\% of the flux at the lower inner target, while the opposite is observed in the USN $(\delta R>0)$. Similarly, Figure \ref{Fig:heat_asymmetry}\textcolor{blue}{b} shows that the up-down asymmetry between outer targets at the LFS increases with the level of magnetic imbalance. The asymmetry is more pronounced in the LSN than in the USN, with a higher heat flux on the lower outer targets. A larger heat flux to the lower outer target is even observed in balanced DN configurations, a phenomenon also attributed to diamagnetic effects, as explained below.

Figure \ref{Fig:heat_asymmetry}\textcolor{blue}{c} and \ref{Fig:heat_asymmetry}\textcolor{blue}{d} reveal that higher resistivity, which in turn enhances turbulent transport, reduces the up-down asymmetry. This reduction is particularly evident at the HFS, where the asymmetry between upper and lower inner targets nearly disappears at high resistivity. On the other hand, the mechanisms driving the heat flux asymmetry are found to be the equilibrium diamagnetic and mean poloidal $E \times B$ drifts, that convect plasma poloidally outside the separatrix (the up-down asymmetry of the perpendicular heat flux crossing the separatrix is smaller and, in general, not directly correlated with the asymmetry at the targets). In addition, the contribution from poloidal $E \times B$, proportional to $\nabla \phi \sim \phi / L_\phi$, is often small compared to the diamagnetic contribution, which is proportional to $\nabla p_e \sim p_e / L_p$, as $L_\phi$ is significantly larger than $L_p$, and the reversal of $L_\phi$ sign between the inner and outer separatrix offsets its global contribution. The importance of the poloidal diamagnetic drift is confirmed by a simulation of the balanced DN with the reverse field. In this configuration, the lower outer target receives a heat flux that is 20\% greater than that directed towards the upper outer target, while the opposite occurs with forward field direction.

These simulations results provide insights into the physical mechanisms driving heat asymmetry in both balanced and unbalanced DN configurations. These include the direction of $B_T$, the magnetic imbalance, and the competition between poloidal diamagnetic drift and radial turbulent transport. Based on these observations, in the following section, we develop an analytical scaling law to qualitatively predict heat asymmetry on the outer targets in DN configurations.

\section{Derivation of the heat asymmetry scaling law}\label{Sec4}
In a DN configuration, the increase in power-sharing asymmetry is often represented by a logistic function \cite{Petrie2001, Brunner2018},
\begin{align}
    \frac{P_{\textrm{LFS,L}}-P_{\textrm{LFS,U}}}{P_{\textrm{LFS,L}}+P_{\textrm{LFS,U}}} = \tanh{\bigg(-\frac{\delta R}{2L_{p}}\bigg)},
\label{scaling_tanh}
\end{align}
assuming an even heat distribution when $\delta R=0$. This is in contrast to observations in both GBS simulations and experiments \cite{Brunner2018, Temmerman2010}. While a substitution $\delta R \rightarrow \delta R - R_*$ to account for an imbalance at $\delta R=0$, with $R_*$ being a correction factor \cite{Brunner2018}, has been considered in the past, the predictive capabilites of the logistic fit for power-sharing in fusion devices are rather limited. This limitation is mainly due to the uncertainties of $R_*$ and $L_p$, which are typically unknown and have to be expressed as a function of the tokamak operational parameters.

In Section \ref{Sec3}, the main reasons behind the heat asymmetry at the outer targets in DN configurations are identified as the magnetic imbalance, and the competition between the poloidal diamagnetic drift, directing the heat flow to either the upper or lower targets, depending on the direction of $B_T$, and the radial turbulent transport. At the same time, the simulations results allow us to neglect the heat load to the inner targets in the development of an analytical scaling law for power load asymmetry at the outer targets. Based on these observations, we propose the following form of the scaling law for the heat asymmetry
\begin{align}
      \lvert q_{\parallel, \textrm{LO}} - q_{\parallel, \textrm{UO}}\rvert
      =q_{\textrm{asym}} &= q_\psi \bigg[\alpha_{g} + (1-\alpha_{g})\alpha_d K \bigg], 
      \label{scaling_heat_asymmetry}
\end{align}
where $q_\psi$ is the total heat flux crossing the separatrix, $\alpha_{g}$ is a geometrical factor accounting for the effect of the magnetic imbalance, $\alpha_d$ is a dimensionless diamagnetic parameter that includes the effects of both drift and turbulence, and $K$ is a numerical value to account for order of magnitude estimate and determined by fitting simulations and experimental results. We now estimate $\alpha_g$ and $\alpha_d$.  In unbalanced DN configurations, the secondary separatrix affects the heat asymmetry through the ratio of the pressure scale length to the magnitude of $\delta R$, a phenomenon that can be represented as
\begin{align}
    \alpha_{g} &= \frac{1}{L_p}\int_0^{|\delta R|}\exp{\bigg( -\frac{x}{2L_p} \bigg)} dx = 1-\exp{\bigg( \Big\lvert -\frac{\delta R}{2L_p}\Big\rvert \bigg)},
\label{geo_factor}
\end{align}
where $\alpha_g$ vanishes for $\delta R=0$ and $\alpha_{g}\rightarrow 1$ when $ \lvert \delta R \rvert \rightarrow +\infty$, approaching the SN configuration (in this case, the entire heat flux is directed towards the lower targets, yielding $q_{\textrm{asym}}=q_\psi$). 

The $\alpha_d$ parameter accounts for the competing effects of the diamagnetic drift and turbulence in generating heat asymmetry. Considering the dominant RBMs turbulence in our simulations, we define it as $\alpha_d=\lvert \bm{v}_d \rvert k_{\textrm{RBM}}/\gamma_{\textrm{RBM}}$ where $\bm{v}_d$ represents the diamagnetic velocity and $t_{\textrm{RBM}}$ and $L_{\textrm{RBM}}$ denote the characteristic time and length of the RBMs, respectively, with $k_{\textrm{RBM}}=1/\sqrt{\bar{n}\nu q^2 \gamma_{\textrm{RBM}}}$ and $\gamma_{\textrm{RBM}}=\sqrt{(2\bar{T}_e)/(\rho_* L_p)}$. We approximate $\lvert \bm{v}_d \rvert \simeq \bar{p}_e/(\bar{n} L_p)$. The larger the value of $\alpha_d$, the more pronounced the asymmetry driving mechanisms become. In normalized units, $\alpha_d$ can be written as
\begin{align}
    \alpha_d \sim \rho_*^{3/4} \bar{n}^{-3/2} \bar{p}_e L_p^{-1/4} \nu_0^{-1/2}q^{-1}.
    \label{diag_analytical}
\end{align}
Stronger turbulence, present at large values of $\nu_0$, leads to a smaller value of $\alpha_d$, which in turn reduces the heat flux asymmetry, in agreement with the findings discussed in Section \ref{Sec3}. We note that the $\alpha_d$ parameter, introduced in Ref.\,\cite{Guzdar1993}, has been used to identify the nature of plasma turbulence. Drift waves (DWs) are expected to be the driving instability when $\alpha_d \geq 1$, and RBMs when $\alpha_d \ll 1$. In all our simulations, the values of $\alpha_d$ are found to be below 0.1, indicating that turbulence is mainly driven by RBMs.  

\begin{figure}[htbp]
\begin{center}	\includegraphics[width=\textwidth]{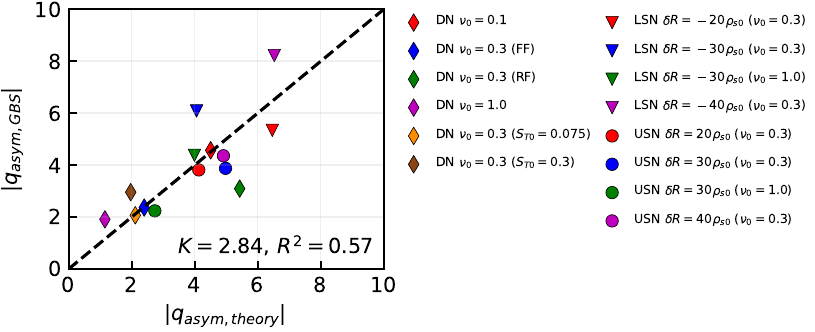}
\caption{Comparison of the heat flux asymmetry between the analytical prediction derived from Eq.\,(\ref{scaling_heat_asymmetry}) and that measured from nonlinear GBS simulations. The prefactor is set to $K=2.84$ for all the simulations, yielding an $R^2$-score of $0.57$.}
\label{qasym_scaling}
\end{center}
\end{figure}

Figure \ref{qasym_scaling} illustrates a comparison of the heat flux asymmetry $q_{\rm{asym}}$ between the analytical scaling law outlined in Eq.\,(\ref{scaling_heat_asymmetry}) and the results from nonlinear GBS simulations. We set the prefactor $K=2.84$, a value determined by using linear regression methods. Overall, the scaling law effectively captures the trend of heat asymmetry observed across various simulations, in which we vary both magnetic imbalances, plasma resistivity, and direction of the magnetic field. Consistent with the findings in Fig.\,\ref{Fig:heat_asymmetry}, enhanced plasma turbulence leads to reduced heat asymmetry between the upper and lower outer divertor targets due to small $\alpha_d$. In contrast, we observe a more pronounced heat asymmetry when turbulent transport is reduced. In this case, the effect of the diamagnetic parameter $\alpha_d$ becomes more significant with respect to turbulence, thereby facilitating that the heat flux channels to either the upper or lower outer targets. Furthermore, we note that heat asymmetry increases in proportion to $\delta R$. As $\delta R$ increases, it amplifies this asymmetry, eventually yielding a SN-like configuration. The quality of the scaling law is quantified by the $R^2$-score, resulting in a value of $R^2=0.57$. This score indicates a reasonable correlation between our analytical predictions and the simulation results, with discrepancies that can be attributed in part to the fact that we neglect the heat flux arriving at the inner targets, and having neglected the effect of the poloidal $E \times B$ flow. 

\section{Validation of power-sharing scaling law}\label{Sec5}
We validate the analytical power-sharing scaling law for outer targets in DN configurations in Eq.\,(\ref{scaling_heat_asymmetry}), using experimental data from TCV \cite{Reimerdes2022}. A recent study on DN configurations in TCV reveals that transitioning from a LSN to DN configuration leads to a reduction of, approximately, 30\% of heat flux at the lower outer targets \cite{Fevrier2021}. In the present study, we analyze five TCV DN L-mode discharges (discharges \#77144, \#77146, \#77153, \#77155, \#77156), which represent the extension of the TCV-X21 plasmas \cite{Oliveira2022} to DN configurations. In these discharges, where the value of $\delta R$ is varied while other parameters are kept relatively constant, two different core densities are considered to explore the effects of density on power sharing. The first two shots (\#77144 and \#77146) and the subsequent three shots (\#77153, \#77155, \#77156) are executed under two different scenarios within the same campaign, the differences in plasma conditions across these discharges providing an ideal validation testbed.

In order to validate our scaling law, the dependence of $\bar{T}_e$ in Eqs.\,(\ref{Lp_analytical}) and (\ref{diag_analytical}) is expressed in terms of $S_p$ and $L_p$. To this end, we integrate the heat flux at the vessel wall over the SOL width and balance it with the input power,
\begin{equation}
    \iint \bar{p}_e \bar{c}_s \bm{b} \cdot d\bm{S} \simeq S_p, 
\end{equation}
where $d\bm{S}$ is the differential target area, under the assumption of a low-recycling regime and plasma flows at the divertor plate approaching sound speed. Assuming that the electron pressure and temperature decrease exponentially within the SOL, with the $L_p$ and $L_T$ scale lengths, where $L_T \simeq (1+\eta_e)L_p/\eta_e$ with $\eta_e \simeq 0.77$ \cite{Ricci2008}, we can express the average electron temperature at the LCFS as:
\begin{align}
    \bar{T}_e \sim \bigg[ \bigg( 1+\frac{\eta_e}{2(1+\eta_e)}\bigg)\frac{S_p}{\bar{n}L_p}\bigg]^{2/3}.
\end{align}
Substituting $\bar{T}_e$ into Eqs.\,(\ref{Lp_analytical}) and (\ref{diag_analytical}), we derive the following expression for $L_p$,
\begin{align}
    L_p &\sim  \bigg[\mathcal{C}(\kappa, \delta, q)^9 \bigg( 1+\frac{\eta_e}{2(1+\eta_e)}\bigg)^{8}\rho_*^3 \bar{n}^{10} S_p^{-4}
     L_\chi^{12}\nu_0^6  q^{12} \bigg]^{1/17}, \label{Lp_anal2}
\end{align}
and for $\alpha_d$,
\begin{align}
    \alpha_d &\sim \bigg( 1+\frac{\eta_e}{2(1+\eta_e)}\bigg)^{2/3} \rho_*^{3/4} \bar{n}^{-7/6} S_p^{2/3}L_p^{-11/12}\nu_0^{-1/2}q^{-1}.
    \label{diag_analytical2}
\end{align}
Finally, we reformulate the scaling law in Eq.\,(\ref{scaling_heat_asymmetry}) using engineering parameters, including the tokamak major and minor radii, $R_0$ and $a$, plasma elongation and triangularity, $\kappa$ and $\delta$, the toroidal magnetic field, $B_T$, safety factor, $q_{95}$, and the power crossing the separatrix, $P_{\textrm{SOL}}$. More precisely, (1) we express the total heat flux $q_\psi$ crossing the separatrix as a function of $S_p$, similarly to Eq.\,(\ref{Lp_analytical}), (2) we replace the injected power source $S_p$ with $P_{\rm{SOL}}/(2\pi R_0)$, and (3) we express the plasma resistivity $\nu_0$ using Eqs.\,(\ref{Collisionality}), (\ref{Lp_anal2}) and (\ref{diag_analytical2}). This yields the scaling law
\begin{align}
    \lvert{P_{\textrm{LO}}-P_{\textrm{UO}}}\rvert =P_{\textrm{asym}} = P_{\textrm{SOL}} \bigg[\alpha_g + (1-\alpha_g)\alpha_d K \bigg],
    \label{Power_load_scaling_law}
\end{align}
where $q_{\textrm{SOL}}$ is now replaced by the power $P_{\textrm{SOL}}$, and the pressure gradient length $L_p$ and the dimensionless parameter $\alpha_d$ are now represented in physical units as follows,
\begin{align}
    L_p &\sim 1.95 \, \mathcal{C}(\kappa, \delta, q)^{9/17} A^{1/17} q^{12/17} R_0^{7/17} P_{\textrm{SOL}}^{-4/17} n^{10/17}B_T^{-12/17}L_\chi^{12/17},\label{Lp_analytical2}
\end{align}
and
\begin{align}
    \alpha_d&\sim 1.14 \, A^{13/12}R_0^{-23/12}B_T^{-1}L_p^{-11/12}P_{\textrm{SOL}}^{2/3}n^{-7/6}q^{-1}. \label{alpha_d2} 
\end{align}
Here, $L_p$ is expressed in mm, $A$ denotes the mass number of the main plasma ions, $P_{\rm{SOL}}$ is measured in MW, $n_e$ represents the edge density at $q_{\rm{95}}$ in units of $10^{19}\rm{m}^{-3}$ and $B_T$ is in T.

\begin{figure*}[htbp]
  \centering
  % First subfigure
  \begin{subfigure}[b]{0.23\textwidth}
    \includegraphics[width=\textwidth]{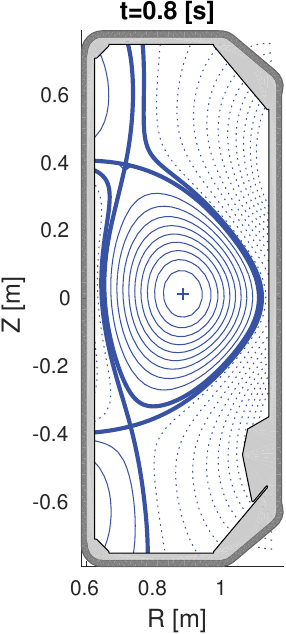}
    \caption{$\delta R=8$ [mm]}
    \label{fig:sub1}
  \end{subfigure}
  % Second subfigure
  \begin{subfigure}[b]{0.23\textwidth}
    \includegraphics[width=\textwidth]{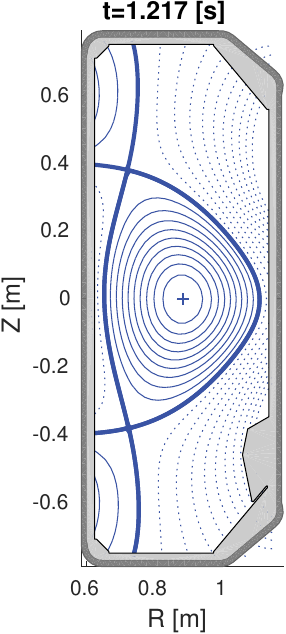}
    \caption{$\delta R=0$}
    \label{fig:sub2}
  \end{subfigure}
  % Third subfigure
  \begin{subfigure}[b]{0.23\textwidth}
    \includegraphics[width=\textwidth]{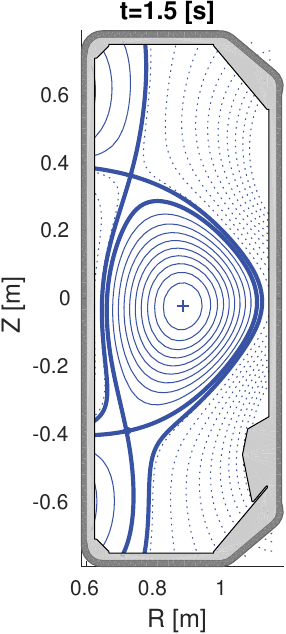}
    \caption{$\delta R=-15$ [mm]}
    \label{fig:sub3}
  \end{subfigure}
  % Fourth subfigure
  \begin{subfigure}[b]{0.23\textwidth}
    \includegraphics[width=\textwidth]{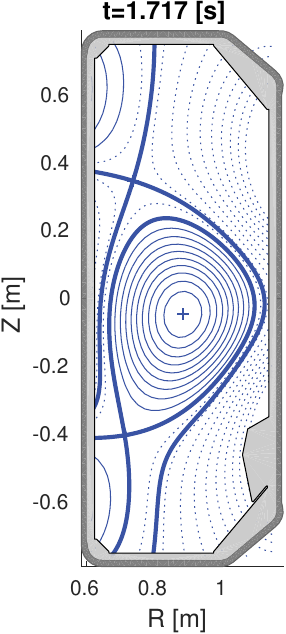}
    \caption{$\delta R=-29$ [mm]}
    \label{fig:sub4}
  \end{subfigure}
  \caption{Time evolution of the DN magnetic configurations during discharge \#77144, with $\delta R$ varying from 8 mm to -29 mm. The configuration transitions from USN to LSN, passing through the balanced DN configuration with $\delta R=0$.}
  \label{TCV_DN_config}
\end{figure*}

During the discharge \#77144, four distinct magnetic configurations that illustrate the transition from USN to LSN are generated, with $\delta R$ varying from 8 mm to -29 mm, as shown in Fig.\,\ref{TCV_DN_config}. The power reaching the outer targets is measured using wall-embedded Langmuir Probes (LPs) \cite{Fevrier2018, Oliveira2019}. Additionally, the $\nabla B$-drift direction is consistently oriented away from the lower X-point.

\begin{figure}[H]
\begin{center}	
\includegraphics[width=0.45\textwidth]{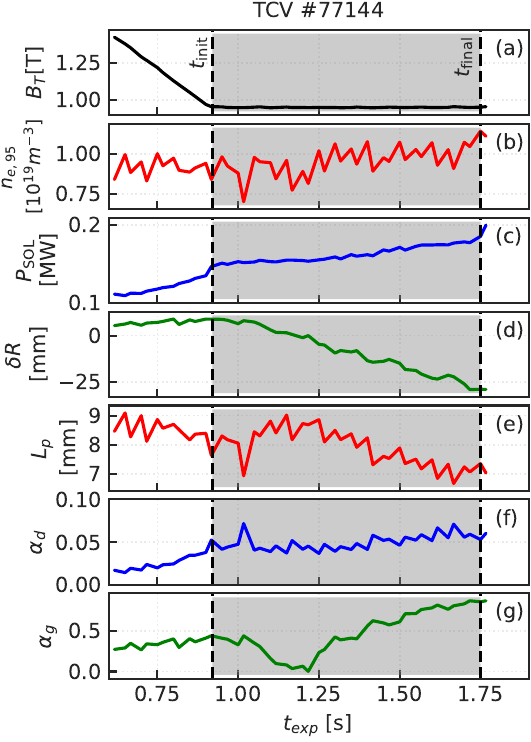}
\caption{Time trace of the main plasma parameters (a--d) and analytical parameters appearing in the scaling law (e--g) during the transition from USN $(\delta R<0)$ to LSN $(\delta R>0)$ from TCV discharge \#77144. We show (a) the toroidal magnetic field $B_T$ [T], (b) edge electron density $n_{e,95}$ [$10^{-19}m^{-3}$], (c) power entering the SOL $P_{\rm{SOL}}$ [MW], (d) inter-separatrix distance $\delta R$ [mm], (e) electron pressure gradient length $L_p$ [mm] from Eq.\,(\ref{Lp_analytical2}), (f) dimensionless diamagnetic parameter $\alpha_d$ from Eq.\,(\ref{alpha_d2}), and (g) geometrical factor $\alpha_g$ from Eq.\,(\ref{geo_factor}). The shaded area between $t_{\rm{init}}$ and $t_{\rm{final}}$ represents the time interval within which the power load asymmetry is validated against the experimental data.}
\label{TCV_Data_trace}
\end{center}
\end{figure}

Figure \ref{TCV_Data_trace} presents a temporal analysis of the plasma parameters throughout the transition from USN to LSN in TCV discharge \#77144. We maintain a low-$B_T$ $(0.9\rm{T})$ scenario and a low line-averaged electron density $\langle n_e\rangle \leq 4 \times 10^{19}\rm{m}^{-3}$ for all five discharges, resulting in attached divertor conditions. It is important to note that measurement uncertainties are important, particularly in the calculation of $P_{\rm{SOL}}=P_{\rm{Ohmic}}-P_{\rm{rad}}$, where $P_{\rm{Ohmic}}$ is the Ohmic heating power and $P_{\rm{rad}}$ is the radiation power, with errors up to 40\%. The discharges are Ohmic-only, and characterized by a plasma current $I_p=160\rm{kA}$. The validation of the TCV power load asymmetry is conducted within the time window from 0.8s to 1.75s, wherein the inter-separatrix distance $\delta R$ varies from 8 mm to -29 mm, resulting in a transition from the USN to LSN configurations.

The parameters derived from the experimental data, namely the electron pressure gradient length $L_p$ in Eq.\,(\ref{Lp_analytical2}), the dimensionless diamagnetic parameter $\alpha_d$ in Eq.\,(\ref{alpha_d2}), and the geometrical factor $\alpha_g$ in Eq.\,(\ref{geo_factor}) are also presented in Fig.\,\ref{TCV_Data_trace} (e--f). These parameters are computed using the experimentally measured values from Fig.\,\ref{TCV_Data_trace} (a--d). Notably, the value of $\alpha_d$ remains below 0.1, suggesting that RBMs dominate over DWs, and consistent with the fact that the considered discharge is an L-mode plasma \cite{Giacomin2022}. Furthermore, variations in $\delta R$ modify the time variation of $\alpha_g$, which affects, in turn, the power load distribution between the upper and lower divertor targets across different DN configurations. 

\begin{figure}
\begin{center}	
\includegraphics[width=0.4\textwidth]{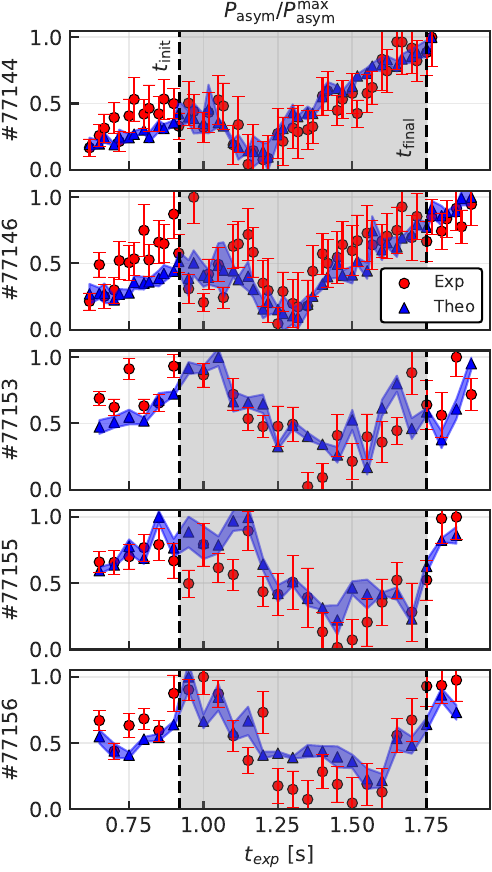}
\caption{Comparative validation results showing the normalized power load asymmetry $(P_{\rm{asym}}/P^{\rm{max}}_{\rm{asym}})$ for five TCV discharges over time. Experimental data (Exp) are denoted by red circles, while theoretical predictions (Theo) are represented by blue triangles. The dashed vertical lines mark the start ($t_{\rm{init}}$) and end ($t_{\rm{final}}$) of the validation time window. We set $K=2.84$ in the scaling law.}
\label{theory_prediction}
\end{center}
\end{figure}

Figure \ref{theory_prediction} shows the validation of the analytical scaling law for power load asymmetry, derived from the engineering parameters detailed in Eq.\,(\ref{Power_load_scaling_law}), against empirical data from the TCV DN discharges. The power load asymmetry, $P_{\rm{asym}}=|P_{\rm{LO}}-P_{\rm{UO}}|$ is computed from the difference in the heat fluxes reaching the lower and upper outer targets. Specifically, $P_{\rm{LO}}$ is calculated as $\int \bm{q}_{\rm{LO}}\cdot d\bm{S}$, representing the heat flux reaching the LO target. Similarly, $P_{\rm{UO}}$ is the heat flux received by the UO target. To maintain consistency with the analysis of the nonlinear simulations, we use the prefactor $K=2.84$ in Fig.\,\ref{qasym_scaling} (see Sec. \ref{Sec4}). In addition, we make use of the normalized power load asymmetry ($P_{\rm{asym}}/P^{\rm{max}}_{\rm{asym}}$), where $P^{\rm{max}}_{\rm{asym}}$ represents the maximum value of power asymmetry observed in each discharge. This normalization allows for a direct comparison between experimental and theoretical values, effectively highlighting the variations in asymmetry and the degree of agreement between theoretical predictions and experimental data. The dashed vertical lines in the figure define the start, $t_{\rm{init}}$, and end, $t_{\rm{final}}$, times of the validation interval.

While the analytical predictions generally correspond well with the experimentally observed power load asymmetry across different TCV-DN discharges, some deviations are observed, with factors of up to two. These discrepancies can be attributed to several factors, for instance, the neglected power load on the inner target and the assumption that the entire heat flux crossing the separatrix flows to the targets, disregarding radial losses due to blobs that reach the vessel walls, and uncertainties in measurements, particularly evident in calculating $P_{\rm{SOL}}$.

\section{Conclusion}\label{Sec6}
In the present study, both balanced and unbalanced DN configurations in L-mode plasmas are investigated using nonlinear, global, flux-driven, two-fluid GBS simulations. The use of DN configurations often exhibits significant power-sharing asymmetry between the upper and lower divertor targets. To qualitatively understand this asymmetry, a series of nonlinear GBS simulations is carried out by varying parameters such as plasma resistivity, the direction of the toroidal field and the magnetic imbalance.

First, the expression of $L_p$ is found to be identical to that of the SN configuration in Ref.\,\cite{Lim2023}, suggesting that introducing a secondary separatrix does not significantly affect the level of background RBM turbulence in the near SOL region and $L_p$ scale length. A comparison between the analytical $L_p$ estimate and numerical values from nonlinear GBS simulations reveals an overall good agreement.

An in-depth analysis of the power-sharing asymmetry in DN configurations identifies several key factors contributing to the observed heat asymmetry. First, the competition between poloidal diamagnetic drift and radial turbulent transport plays a crucial role in determining the power-sharing asymmetry between the upper and the lower outer targets. Additionally, magnetic imbalance $\delta R$ affects the power-sharing asymmetry, showing a non-negligible level of asymmetry even in the balanced DN configurations. This asymmetry tends to increase with respect to the value of $\delta R$ as it approaches a SN-like configuration. 

Based on these observations, an analytical scaling law to predict the power-sharing asymmetry observed in GBS simulations is derived, introducing a dimensionless $\alpha_d$ parameter to capture the relative importance of drifts and turbulence, along with a geometrical factor $\alpha_g$, to account for the magnetic imbalance. When compared with numerical results, the analytical scaling law reflects the overall trend with an $R^2$-score of $0.5$. 

To validate the scaling law against experimental data, we analyze five DN discharges in TCV L-mode plasmas, carried out to study the variations in magnetic imbalance $\delta R$. For this purpose, we reformulate the analytical scaling law in terms of engineering parameters, finding a reasonable agreement between theoretical predictions and experimental observations. The comparison is affected by measurement uncertainties, particularly in $P_{\rm{SOL}}$, and our predictive capabilities might be reduced due to the simplifying assumptions used in deriving the scaling law, including the neglected power load to the inner targets.

To enhance the reliability of the scaling law for predicting power load asymmetry, a more comprehensive validation is necessary. This includes the analysis of multi-machine datasets to ensure wider applicability and reliability. In particular, since DN configurations with negative triangularity (NT) plasmas could potentially offer enhanced advantages in power handling within fusion devices, the scaling law developed in this study, which accounts for plasma shaping effects, should be validated in this configuration to accurately predict the heat asymmetry.
\begin{acknowledgments}
This work has been carried out within the framework of the EUROfusion Consortium, partially funded by the European Union via the Euratom Research and Training Programme (Grant Agreement No 101052200 — EUROfusion). The Swiss contribution to this work has been funded by the Swiss State Secretariat for Education, Research and Innovation (SERI). Views and opinions expressed are however those of the author(s) only and do not necessarily reflect those of the European Union, the European Commission or SERI. Neither the European Union nor the European Commission nor SERI can be held responsible for them. This work was supported in part by the Swiss National Science Foundation. The simulations presented herein were carried out in part on the CINECA Marconi supercomputer under the TSVV-1 and TSVV-2 projects and in part at the Swiss National Supercomputing Center (CSCS) under the s1177 and s1241 projects.
\end{acknowledgments}
%\appendix
%\input{8_Appendix}
\newpage
%\bibliographystyle{abbrv}
%\bibliographystyle{ieetr} 
% ieetr -> Ordering by appearance
\bibliographystyle{unsrt}
\bibliography{9_bib.bib}

\begin{thebibliography}{10}

\bibitem{Donne2019}
A.~J.~H. Donné.
\newblock The european roadmap towards fusion electricity.
\newblock {\em Philos. Trans. Royal Soc.},
  \href{https://dx.doi.org/10.1098/rsta.2017.0432} {\textbf{377} 20170432},
  (2019).

\bibitem{Soukhanovskii2017}
V.~A. Soukhanovskii.
\newblock A review of radiative detachment studies in tokamak advanced magnetic
  divertor configurations.
\newblock {\em Plasma Phys. Control. Fusion},
  \href{https://dx.doi.org/10.1088/1361-6587/aa6959} {\textbf{59} 064005},
  (2017).

\bibitem{Wenninger2018}
R.~Wenninger et~al.
\newblock {Power Handling and Plasma Protection Aspects that affect the design
  of the DEMO divertor and first wall}.
\newblock {\em (IAEA-CN--234) IAEA Conference Proceedings}, (2018).

\bibitem{Reimerdes2020}
H.~Reimerdes et~al.
\newblock Assessment of alternative divertor configurations as an exhaust
  solution for {DEMO}.
\newblock {\em Nucl. Fusion},
  \href{https://doi.org/10.1088/1741-4326/ab8a6a}{\textbf{60} 066030}, (2020).

\bibitem{Militello2021}
F.~Militello et~al.
\newblock {Preliminary analysis of alternative divertors for DEMO}.
\newblock {\em Nucl. Mater. Energy},
  \href{https://doi.org/10.1016/j.nme.2021.100908}{\textbf{26} 100908}, (2021).

\bibitem{Smick2013}
N.~Smick, B.~LaBombard, and I.H. Hutchinson.
\newblock {Transport and drift-driven plasma flow components in the Alcator
  C-Mod boundary plasma}.
\newblock {\em Nucl. Fusion},
  \href{https://doi.org/10.1088/0029-5515/53/2/023001}{\textbf{53} 023001},
  (2013).

\bibitem{LaBombard2017}
B.~LaBombard et~al.
\newblock High-field side scrape-off layer investigation: Plasma profiles and
  impurity screening behavior in near-double-null configurations.
\newblock {\em Nucl. Mater. Energy},
  \href{https://doi.org/10.1016/j.nme.2016.10.006}{{12} 139}, (2017).

\bibitem{LaBombard2017_2}
B.~LaBombard et~al.
\newblock Impurity screening behavior of the high-field side scrape-off layer
  in near-double-null configurations: prospect for mitigating
  plasma{\textendash}material interactions on {RF} actuators and first-wall
  components.
\newblock {\em Nucl. Fusion},
  \href{https://doi.org/10.1088/1741-4326/aa6dd2}{{57} 076021}, (2017).

\bibitem{Fevrier2021}
O.~F\'evrier et~al.
\newblock {Detachment in conventional and advanced double-null plasmas in TCV}.
\newblock {\em Nucl. Fusion},
  \href{https://doi.org/10.1088/1741-4326/ac27c6}{\textbf{61} 116064}, (2021).

\bibitem{Kim2023}
Boseong Kim et~al.
\newblock {Investigation of performance enhancement by balanced double-null
  shaping in KSTAR}.
\newblock {\em Nucl. Fusion},
  \href{https://dx.doi.org/10.1088/1741-4326/acf677}{\textbf{63} 126013},
  (2023).

\bibitem{Brunner2018}
D.~Brunner, A.Q. Kuang, B.~LaBombard, and J.L. Terry.
\newblock {The dependence of divertor power sharing on magnetic flux balance in
  near double-null configurations on Alcator C-Mod}.
\newblock {\em Nucl. Fusion},
  \href{https://doi.org/10.1088/1741-4326/aac006}{\textbf{58} 076010}, (2018).

\bibitem{Petrie2001}
T.W. Petrie et~al.
\newblock {The effect of divertor magnetic balance on H-mode performance in
  DIII-D}.
\newblock {\em J. Nucl. Mater.},
  \href{https://doi.org/10.1016/S0022-3115(00)00492-X}{\textbf{290} 935},
  (2001).

\bibitem{Guo2011}
H.Y. Guo et~al.
\newblock {Recent progress on divertor operations in EAST}.
\newblock {\em J. Nucl. Mater.},
  \href{https://doi.org/10.1016/j.jnucmat.2010.11.048}{\textbf{415} S369},
  (2011).

\bibitem{Temmerman2011}
G.~{De Temmerman}, A.~Kirk, E.~Nardon, and P.~Tamain.
\newblock Heat load asymmetries in {MAST}.
\newblock {\em J. Nucl. Mater.},
  \href{https://doi.org/10.1016/j.jnucmat.2010.10.003}{\textbf{415} S383},
  (2011).

\bibitem{Morel1999}
K.M Morel, G.F Counsell, and P~Helander.
\newblock {Asymmetries in the divertor power loading in START}.
\newblock {\em J. Nucl. Mater.},
  \href{https://doi.org/10.1016/S0022-3115(98)00851-4}{\textbf{266} 1040},
  (1999).

\bibitem{Contessa2019}
G.M. Contessa et~al.
\newblock {\em {DTT-Divertor Tokamak Test Facility-Interim Design Report}}.
\newblock ENEA, (2019).

\bibitem{Im2016}
Kihak Im, Sungjin Kwon, and Jong~Sung Park.
\newblock {A Preliminary Development of the K-DEMO Divertor Concept}.
\newblock {\em IEEE Trans. Plasma Sci.},
  \href{https://doi.org/10.1109/TPS.2016.2604408}{\textbf{44} 2493}, (2016).

\bibitem{Kuang2020}
A.~Q. Kuang, S.~Ballinger, D.~Brunner, J.~Canik, A.~J. Creely, T.~Gray,
  M.~Greenwald, J.~W. Hughes, J.~Irby, B.~LaBombard, and et~al.
\newblock {Divertor heat flux challenge and mitigation in SPARC}.
\newblock {\em J. Plasma Physics},
  \href{https://dx.doi.org/10.1017/S0022377820001117} {\textbf{86} 865860505},
  (2020).

\bibitem{Wilson2020}
H.~Willson et~al.
\newblock {\em {STEP—on the pathway to fusion commercialization}}.
\newblock 2053-2563. IOP Publishing, (2020).

\bibitem{Marchand1995}
R.~Marchand, M.~Dumberry, Y.~Demers, C.~Cote, G.~Le Clair, J.-M. Larsen,
  X.~Bonnin, and B.J. Braams.
\newblock Up-down symmetry in double null divertor experiments and magnetic
  field topology.
\newblock {\em Nucl. Fusion},
  \href{https://dx.doi.org/10.1088/0029-5515/35/3/I04}{\textbf{35} 297},
  (1995).

\bibitem{Temmerman2010}
G.~De. Temmerman et~al.
\newblock Thermographic study of heat load asymmetries during {MAST} {L}-mode
  discharges.
\newblock {\em Plasma Phys. Control. Fusion},
  \href{https://doi.org/10.1088/0741-3335/52/9/095005}{{52} 095005}, (2010).

\bibitem{Petrie2003}
T.W. Petrie et~al.
\newblock {Changes in edge and scrape-off layer plasma behavior due to
  variation in magnetic balance in DIII-D}.
\newblock {\em J. Nucl. Mater.},
  \href{https://doi.org/10.1016/S0022-3115(02)01459-9}{{313} 834}, (2003).

\bibitem{Petrie2006}
T.W Petrie et~al.
\newblock Variation of particle exhaust with changes in divertor magnetic
  balance.
\newblock {\em Nucl. Fusion},
  \href{https://doi.org/10.1088/0029-5515/46/1/007}{{46} 57}, (2006).

\bibitem{Liu2012}
S.~C. Liu et~al.
\newblock {Divertor asymmetry and scrape-off layer flow in various divertor
  configurations in Experimental Advanced Superconducting Tokamak}.
\newblock {\em Phys. Plasmas},
  \href{https://doi.org/10.1063/1.4707396}{\textbf{19} 042505}, (2012).

\bibitem{Rognlien1999}
T.D. Rognlien, G.D. Porter, and D.D. Ryutov.
\newblock {Influence of $E \times B$ and $\nabla B$ drift terms in 2D edge/SOL
  transport simulations}.
\newblock {\em J. Nucl. Mater.},
  \href{https://doi.org/10.1016/S0022-3115(98)00835-6}{\textbf{266} 654},
  (1999).

\bibitem{Cohen1999}
R.~H. Cohen and D.~Ryutov.
\newblock Drifts, boundary conditions and convection on open field lines.
\newblock {\em Phys. Plasmas},
  \href{https://doi.org/10.1063/1.873455}{\textbf{6} 1995}, (1999).

\bibitem{Rensink2000}
M.E. Rensink, S.L. Allen, G.D. Porter, and T.D. Rognlien.
\newblock {Simulation of Double-Null Divertor Plasmas with the UEDGE Code}.
\newblock {\em Contrib. Plasma. Phys},
  \href{https://doi.org/10.1002/1521-3986(200006)40:3/4<302::AID-CTPP302>3.0.CO;2-L}{\textbf{40}
  302-308}, (2000).

\bibitem{Rubino2020}
G.~Rubino et~al.
\newblock {Effect of drifts on SOL plasma in DTT Double Null configuration}.
\newblock {\em 47th EPS Conference on Plasma Physics},
  \href{http://ocs.ciemat.es/EPS2021PAP/pdf/P3.1026.pdf}{\textbf{P3} 1026},
  (2020).

\bibitem{Du2015}
H.~Du et~al.
\newblock {Effects of drifts and ballooning instability on the divertor
  in–out asymmetry in EAST tokamak}.
\newblock {\em J. Nucl. Materials},
  \href{https://doi.org/10.1016/j.jnucmat.2014.12.093}{\textbf{463} 485},
  (2015).

\bibitem{Schaffer1997}
M.J. Schaffer et~al.
\newblock {Pfirsch-Schluter currents in the JET divertor}.
\newblock {\em Nucl. Fusion},
  \href{https://dx.doi.org/10.1088/0029-5515/37/1/I07}{\textbf{37} 83}, (1997).

\bibitem{Asakura2004}
N.~Asakura et~al.
\newblock {Driving mechanism of SOL plasma flow and effects on the divertor
  performance in {JT}-60U}.
\newblock {\em Nucl. Fusion},
  \href{https://doi.org/10.1088/0029-5515/44/4/004}{\textbf{44} 503}, (2004).

\bibitem{Osawa2023}
R.T. Osawa, D.~Moulton, S.L. Newton, S.S. Henderson, B.~Lipschultz, and
  A.~Hudoba.
\newblock {SOLPS-ITER analysis of a proposed STEP double null geometry: impact
  of the degree of disconnection on power-sharing}.
\newblock {\em Nucl. Fusion},
  \href{https://dx.doi.org/10.1088/1741-4326/acd863}{\textbf{63} 076032},
  (2023).

\bibitem{Petrie1999}
T.W. Petrie et~al.
\newblock {A comparison of plasma performance between single-null and
  double-null configurations during Elming H-mode}.
\newblock {\em EPS 1999}, \href{https://www.osti.gov/servlets/purl/766942}{
  GA-A23155}, (1999).

\bibitem{Maggi2014}
C.F. Maggi et~al.
\newblock {L{\textendash}H power threshold studies in {JET} with Be/W and C
  wall}.
\newblock {\em Nucl. Fusion},
  \href{https://doi.org/10.1088/0029-5515/54/2/023007}{\textbf{54} 023007},
  (2014).

\bibitem{Beadle2020}
C.~Beadle and P~Ricci.
\newblock Understanding the turbulent mechanisms setting the density decay
  length in the tokamak scrape-off layer.
\newblock {\em J. Plasma Phys.},
  \href{https://doi.org/10.1017/S0022377820000094}{\textbf{86} 175860101},
  (2020).

\bibitem{Ahomantila2021}
L.~Aho-Mantila et~al.
\newblock Scoping the characteristics and benefits of a connected double-null
  configuration for power exhaust in eu-demo.
\newblock {\em Nucl. Mater. Energy},
  \href{https://doi.org/10.1016/j.nme.2020.100886}{\textbf{26} 100886}, (2021).

\bibitem{Innocente2021}
P.~Innocente, L.~Balbinot, H.~Bufferand, and G.~Ciraolo.
\newblock Study of the double null divertor configuration in dtt.
\newblock {\em Nucl. Mater. Energy},
  \href{https://doi.org/10.1016/j.nme.2021.100985}{{27} 100985}, (2021).

\bibitem{Porter2010}
G.~D. Porter, T.~W. Petrie, T.~D. Rognlien, and M.~E. Rensink.
\newblock {UEDGE simulation of edge plasmas in DIII-D double null
  configurations}.
\newblock {\em Phys. Plasmas},
  \href{https://doi.org/10.1063/1.3499666}{\textbf{17} 112501}, (2010).

\bibitem{Reimerdes2022}
H.~Reimerdes et~al.
\newblock {Overview of the TCV tokamak experimental programme}.
\newblock {\em Nucl. Fusion},
  \href{https://dx.doi.org/10.1088/1741-4326/ac369b}{\textbf{62} 042018},
  (2022).

\bibitem{Zeiler1997}
A.~Zeiler et~al.
\newblock {Nonlinear reduced Braginskii equations with ion thermal dynamics in
  toroidal plasma}.
\newblock {\em Phys. Plasams}, \href{https://doi.org/10.1063/1.872368}
  {\textbf{4} 2134}, (1997).

\bibitem{Ricci2012}
P.~Ricci, F.~D. Halpern, S.~Jolliet, J.~Loizu, A.~Mosetto, A.~Fasoli, I.~Furno,
  and C.~Theiler.
\newblock Simulation of plasma turbulence in scrape-off layer conditions: the
  gbs code, simulation results and code validation.
\newblock {\em Plasma Phys. Control. Fusion},
  \href{https://dx.doi.org/10.1088/0741-3335/54/12/124047}{\textbf{54} 124047},
  (2012).

\bibitem{Giacomin2021}
M.~Giacomin et~al.
\newblock {The GBS code for the self-consistent simulation of plasma turbulence
  and kinetic neutral dynamics in the tokamak boundary}.
\newblock {\em J. Comput. Phys.},
  \href{https://doi.org/10.1016/j.jcp.2022.111294}{\textbf{463} 111294},
  (2022).

\bibitem{Giacomin2020}
M.~Giacomin and P.~Ricci.
\newblock {Investigation of turbulent transport regimes in the tokamak edge by
  using two-fluid simulations}.
\newblock {\em J. Plasma Phys.},
  \href{https://doi:10.1017/S0022377820000914}{\textbf{5} 86}, (2020).

\bibitem{Riva2017}
F.~Riva et~al.
\newblock Plasma shaping effects on tokamak scrape-off layer turbulence.
\newblock {\em Plasma Phys. Control. Fusion},
  \href{https://doi.org/10.1088/1361-6587/aa5322}{\textbf{59} 035001}, (2017).

\bibitem{Lim2023}
K.~Lim, M.~Giacomin, P.~Ricci, A.~Coelho, O.~Février, D.~Mancini, D.~Silvagni,
  and L.~Stenger.
\newblock {Effect of triangularity on plasma turbulence and the SOL-width
  scaling in L-mode diverted tokamak configurations}.
\newblock {\em Plasma Phys. Control. Fusion},
  \href{https://dx.doi.org/10.1088/1361-6587/acdc52}{\textbf{65} 085006},
  (2023).

\bibitem{Coelho2022}
A.~Coelho, J.~Loizu, P.~Ricci, and M.~Giacomin.
\newblock Global fluid simulation of plasma turbulence in a stellarator with an
  island divertor.
\newblock {\em Nucl. Fusion}, (2022).

\bibitem{Loizu2012}
J.~Loizu, P.~Ricci, F.~D. Halpern, and S.~Jolliet.
\newblock {Boundary conditions for plasma fluid models at the magnetic
  presheath entrance}.
\newblock {\em Phys. Plasmas},
  \href{https://doi.org/10.1063/1.4771573}{\textbf{19} 122307}, (2012).

\bibitem{Mosetto2013}
A.~Mosetto et~al.
\newblock Turbulent regimes in the tokamak scrape-off layer.
\newblock {\em Phys. Plasmas},
  \href{https://doi.org/10.1063/1.4821597}{\textbf{20} 092308}, (2013).

\bibitem{Silvagni2020}
D.~Silvagni et~al.
\newblock {Scrape-off layer ({SOL}) power width scaling and correlation between
  {SOL} and pedestal gradients across L, I and H-mode plasmas at {ASDEX}
  Upgrade}.
\newblock {\em Plasma Phys. Control. Fusion},
  \href{https://doi.org/10.1088/1361-6587/ab74e8}{\textbf{62} 045015}, (2020).

\bibitem{Giacomin2021_2}
M.~Giacomin et~al.
\newblock {Theory-based scaling laws of near and far scrape-off layer widths in
  single-null L-mode discharges}.
\newblock {\em Nucl. Fusion},
  \href{https://doi.org/10.1088/1741-4326/abf8f6}{\textbf{61} 76002}, (2021).

\bibitem{Ricci2008}
P.~Ricci, B.~N. Rogers, and S.~Brunner.
\newblock High- and low-confinement modes in simple magnetized toroidal
  plasmas.
\newblock {\em Phys. Rev. Lett.},
  \href{https://doi.org/10.1103/PhysRevLett.100.225002}{\textbf{100} 225002},
  (2008).

\bibitem{Ricci2013}
P.~Ricci and B.~N. Rogers.
\newblock Plasma turbulence in the scrape-off layer of tokamak devices.
\newblock {\em Phys. Plasmas},
  \href{https://doi.org/10.1063/1.4789551}{\textbf{20} 010702}, (2013).

\bibitem{Rognlien1999_2}
T.~D. Rognlien, D.~D. Ryutov, N.~Mattor, and G.~D. Porter.
\newblock {Two-dimensional electric fields and drifts near the magnetic
  separatrix in divertor tokamaks}.
\newblock {\em Phys. Plasmas},
  \href{https://doi.org/10.1063/1.873488}{\textbf{6} 1851}, (1999).

\bibitem{Chankin2015}
A.~V. Chankin, G.~Corrigan, M.~Groth, P.~C. Stangeby, and JET contributors.
\newblock {Influence of the $E \times B$ drift in high recycling divertors on
  target asymmetries}.
\newblock {\em Plasma Phys. Control. Fusion},
  \href{https://doi.org/10.1088/0741-3335/57/9/095002}{\textbf{57} 095002},
  (2015).

\bibitem{Christen2017}
N~Christen, C~Theiler, TD~Rognlien, ME~Rensink, H~Reimerdes, R~Maurizio, and
  B~Labit.
\newblock {Exploring drift effects in TCV single-null plasmas with the UEDGE
  code}.
\newblock {\em Plasma Phys. Control. Fusion},
  \href{https://dx.doi.org/10.1088/1361-6587/aa7c8e}{\textbf{59} 105004},
  (2017).

\bibitem{Stangeby1996}
P.C Stangeby and A.V Chankin.
\newblock {Simple models for the radial and poloidal E {\texttimes} B drifts in
  the scrape-off layer of a divertor tokamak: Effects on in/out asymmetries}.
\newblock {\em Nucl. Fusion},
  \href{https://doi.org/10.1088/0029-5515/36/7/i02}{\textbf{36} 839}, (1996).

\bibitem{Asakura2000}
N.~Asakura et~al.
\newblock {Measurement of Natural Plasma Flow along the Field Lines in the
  Scrape-Off Layer on the JT-60U Divertor Tokamak}.
\newblock {\em Phys. Rev. Lett.},
  \href{https://10.1103/PhysRevLett.84.3093}{\textbf{84} 3093}, (2000).

\bibitem{Guzdar1993}
P.~N. Guzdar, J.~F. Drake, D.~McCarthy, A.~B. Hassam, and C.~S. Liu.
\newblock {Three‐dimensional fluid simulations of the nonlinear
  drift‐resistive ballooning modes in tokamak edge plasmas}.
\newblock {\em Phys. Fluids B},
  \href{https://doi.org/10.1063/1.860842}{\textbf{5} 3712-3727}, (1993).

\bibitem{Oliveira2022}
D.S. Oliveira, T.~Body, D.~Galassi, C.~Theiler, E.~Laribi, P.~Tamain,
  A.~Stegmeir, M.~Giacomin, W.~Zholobenko, P.~Ricci, H.~Bufferand, J.A. Boedo,
  G.~Ciraolo, C.~Colandrea, D.~Coster, H.~de~Oliveira, G.~Fourestey, S.~Gorno,
  F.~Imbeaux, F.~Jenko, V.~Naulin, N.~Offeddu, H.~Reimerdes, E.~Serre, C.K.
  Tsui, N.~Varini, N.~Vianello, M.~Wiesenberger, C.~Wüthrich, and the
  TCV~Team.
\newblock {Validation of edge turbulence codes against the TCV-X21 diverted
  L-mode reference case}.
\newblock {\em Nucl. Fusion},
  \href{https://dx.doi.org/10.1088/1741-4326/ac4cde}{\textbf{62} 096001},
  (2022).

\bibitem{Fevrier2018}
O.~Février, C.~Theiler, H.~De~Oliveira, B.~Labit, N.~Fedorczak, and
  A.~Baillod.
\newblock {Analysis of wall-embedded Langmuir probe signals in different
  conditions on the Tokamak à Configuration Variable}.
\newblock {\em Rev. Sci. Instrum.},
  \href{https://dx.doi.org/10.1063/1.5022459}{\textbf{89} 053502}, (2018).

\bibitem{Oliveira2019}
H.~De~Oliveira, P.~Marmillod, C.~Theiler, R.~Chavan, O.~Février, B.~Labit,
  P.~Lavanchy, B.~Marlétaz, R.~A. Pitts, and TCV team.
\newblock {Langmuir probe electronics upgrade on the tokamak à configuration
  variable}.
\newblock {\em Rev. Sci. Instrum.},
  \href{https://doi.org/10.1063/1.5108876}{\textbf{90} 083502}, (2019).

\bibitem{Giacomin2022}
M.~Giacomin and P.~Ricci.
\newblock {Turbulent transport regimes in the tokamak boundary and operational
  limits}.
\newblock {\em Phys. Plasmas},
  \href{https://doi.org/10.1063/5.0090541}{\textbf{29} 062303}, (2022).

\end{thebibliography}
\end{document}